\documentclass[manuscript]{aastex}
\usepackage{spr-astr-addons}

\usepackage{hyperref}
\usepackage{multirow}
\usepackage{multicol}
\usepackage{color}

\shorttitle{Analysis of the observed and intrinsic durations of gamma-ray bursts with known redshift}
\shortauthors{M. Tarnopolski}

\begin{document}

\title{Analysis of the observed and intrinsic durations of gamma-ray bursts with known redshift}

\author{M. Tarnopolski}
\affil{Astronomical Observatory of the Jagiellonian University\\ul. Orla 171, 30-244 Krak\'ow, Poland}
\email{mariusz.tarnopolski@uj.edu.pl}

\begin{abstract}
The duration distribution of 408 GRBs with measured both duration $T_{90}$ and redshift $z$ is examined. Mixtures of a number of distributions (standard normal, skew-normal, sinh-arcsinh, and alpha-skew-normal) are fitted to the observed and intrinsic durations using the maximum log-likelihood method. The best fit is chosen via the Akaike information critetion. The aim of this work is to assess the presence of the presumed intermediate GRB class, and to provide a phenomenological model more appropriate than the common mixture of standard Gaussians. While $\log T^{obs}_{90}$ are well described by a truly trimodal fit, after moving to the rest frame the statistically most significant fit is unimodal. To trace the source of this discrepancy, 334 GRBs observed only by {\it Swift}/BAT are examined in the same way. In the observer frame, this results in a number of statistically plausible descriptions, being uni- and bimodal, and with the number of components ranging from one to three. After moving to the rest frame, no unambiguous conclusions may be put forward. It is concluded that the size of the sample is not big enough to infer reliably GRB properties based on a univariate statistical reasoning only.
\end{abstract}

\section{Introduction}\label{intro}
Gamma-ray bursts (GRBs) are the most powerful explosions known in the Universe, with an emission peak in the 200--500 keV region, and the total isotropic energy released of the order $10^{51}$--$10^{54}$ ergs \citep[for recent reviews, see][]{nakar,zhang,gehrels,berger,meszaros15}. They are also one of the most distant astronomical objects discovered, with the highest known redshift of $z\sim 9.4$ measured for GRB090429B \citep{cucchiara}. \citet{mazets} first pointed out hints for a bimodal distribution of $T_b$ (taken to be the time interval within which fall $80-90\%$ of the measured GRB's intensity) drawn for 143 events detected in the KONUS experiment. \citet{kouve} also found a bimodal structure in the $\log T_{90}$ distribution of 222 events from {\it CGRO}/BATSE, based on which GRBs are commonly divided into short ($T_{90}<2\,{\rm s}$) and long ($T_{90}>2\,{\rm s}$) classes, where $T_{90}$ is the time interval from 5\% to 95\% of the accumulated fluence. While generally short GRBs are of merger origin \citep{nakar} and long ones come from collapsars \citep{woosley}, this classification is imperfect due to a large overlap in duration distributions of the two populations \citep{lu,bromberg,bromberg2,shahmoradi,shahmoradi2,Tarnopolski3}. \citet{horvath98} and \citet{mukh} independently discovered a third peak in the duration distribution in the BATSE 3B catalog, located between the short and long groups, and the statistical existence of this intermediate class was claimed to be supported \citep{horvath02} with the use of BATSE 4B data. Interestingly, using clustering techniques, \citet{chatto} established the optimal number of classes to be three, too. Also in {\it Swift}/BAT data evidence for a third component in $\log T_{90}$ was announced \citep{horvath08,zhang2,huja,horvath10,zitouni}. Other datasets, i.e. RHESSI \citep{ripa} and BeppoSAX \citep{horvath09}, are both in agreement with earlier results regarding the bimodal distribution, and the detection of a third component was established on a lower, compared to BATSE and {\it Swift}, significance level. Thence, four different satellites provided hints about the existence of a third class of GRBs.

Those conclusions were based on the finding that a mixture of three standard Gaussians \mbox{(a 3-G)} is a better fit than a mixture of two Gaussians (a 2-G). This is not surprising, because adding parameters to a nested model always results in a better fit (in the sense of a lower $\chi^2$ or a higher maximum log-likelihood $\mathcal{L}$) due to more freedom given to the model to follow the data. The important questions are whether this improvement is statistically significant, can the three components be related to physically distinct classes, and whether the model is an appropriate one---is there a model that is a better fit? (See \citealt{Tarnopolski,Tarnopolski2} for a discussion.) However, even quantifying the relative improvement via \mbox{$p$-values}\footnote{If one has two fits with $\chi^2(\nu_1)$ and $\chi^2(\nu_2)$, then their difference, $\Delta\chi^2$, is distributed like $\chi^2(\Delta\nu)$, where $\Delta\nu$ is the difference in the degrees of freedom (see Appendix A in \citealt{band}, and \citealt{horvath98}). Alternatively, if one uses the log-likelihood to assess the goodness of fit, then twice their difference, $2(\mathcal{L}_1-\mathcal{L}_2)$, is distributed like $\chi^2(\Delta\nu)$. If a $p$-value associated with either of the two versions of $\chi^2(\Delta\nu)$ does not exceed the significance level $\alpha$, one of the fits (with lower $\chi^2$ or higher $\mathcal{L}$) is statistically better than the other \citep{horvath02}. It is crucial to note that these methods may be applied to nested models only.} is not a definite detection of another physical class of astronomical objects. All of the post-BATSE 3B fits were bimodal, not trimodal, even if comprised of three components. The third peak in the BATSE 3B sample \citep{horvath98} was smeared out with the BATSE 4B catalog when more data was gathered (see Fig.~5 in \citealt{zitouni}). It was suggested by \citet{zitouni} that the duration distribution corresponding to the collapsar scenario might not necessarily be symmetrical because of a non-symmetrical distribution of envelope masses of the progenitors. Specifically, it was shown by \citet{Tarnopolski} that the $\log T_{90}$ distribution of GRBs detected by {\it Fermi} is also bimodal for several binnings. Moreover, a number of intrinsically skewed distributions were fitted to the data of BATSE, {\it Swift} and {\it Fermi} \citep{Tarnopolski2}, and it was found that mixtures of two skewed components follow the data at least as good (BATSE and {\it Swift}), or better ({\it Fermi}) than a conventionally used 3-G, and that they are bimodal as well (in the sense of having two local maxima; \citealt{schilling}). Generally, $n$-modality is commonly associated with $n$ populations underlying a distribution. Hence, the existence of an intermediate GRB class is unlikely.

The analysis of the observed durations was performed by many authors, as reviewed above. However, the intrinsic duration---the one in the rest frame---of a GRB is affected by its cosmological distance, and is shorter than the observed one:
\begin{equation}
T^{\rm int}_{90}=\frac{T^{\rm obs}_{90}}{1+z}.
\label{eq0}
\end{equation}
Considering the median redshift of long GRBs, $\tilde{z}_{\rm long}\approx 2$, it is evident that GRBs with $T^{\rm obs}_{90}\lesssim 6\,{\rm s}$ have an intrinsic duration generally smaller than $2\,{\rm s}$, which makes them short ones. Note that the classification of short GRBs is the same in both the observer and rest frames. The analysis of the $T^{\rm int}_{90}$ distribution was performed rarely due to a small number of GRBs with measured redshift: \citet{zhang2} examined 95, \citet{huja} analyzed 130, and \citet{zitouni} investigated 248 {\it Swift} GRBs. While \citet{zhang2} focused on the apparent bimodality, and \citet{huja} did not translate the observed durations to the rest frame, \citet{zitouni} found that a 3-G follows the {\it Swift} data better than a 2-G (in observer as well as in the rest frame; see their Figs.~6 and 7). However, in both frames the distributions were bimodal, yet apparently skewed, and hence the existence of an intermediate class is still unlikely. The plausible explanation of this phenomenon is that there are two GRB classes with intrinsically non-symmetrical duration distributions.

The aim of this article is to perform a statistical analysis of the GRBs with measured redshift in order to test against the existence of the intermediate GRB class. Mixtures of various distributions (standard Gaussians, skew-normal, sinh-arcsinh and alpha-skew-normal) are applied to verify whether the statistical significance of a three-Gaussian fit might by challenged by a mixture of skewed distributions with only two components. Both the observed and intrinsic durations are examined.

This article is organized as follows. In Sect.~\ref{meth} the dataset, fitting methods and the properties of the examined distributions are described as outlined by \citet{Tarnopolski2}. In Sect.~\ref{res1} the study of the sample of all GRBs with measured redshift is presented. This is followed by an analysis of {\it Swift} GRBs with known redshift in Sect.~\ref{res2}. Section~\ref{disc} is devoted to discussion, and in Sect.~\ref{conc} concluding remarks are given.

\section{Data and methods}\label{meth}
\subsection{Dataset}\label{data}
\begin{figure}
\includegraphics[width=\columnwidth]{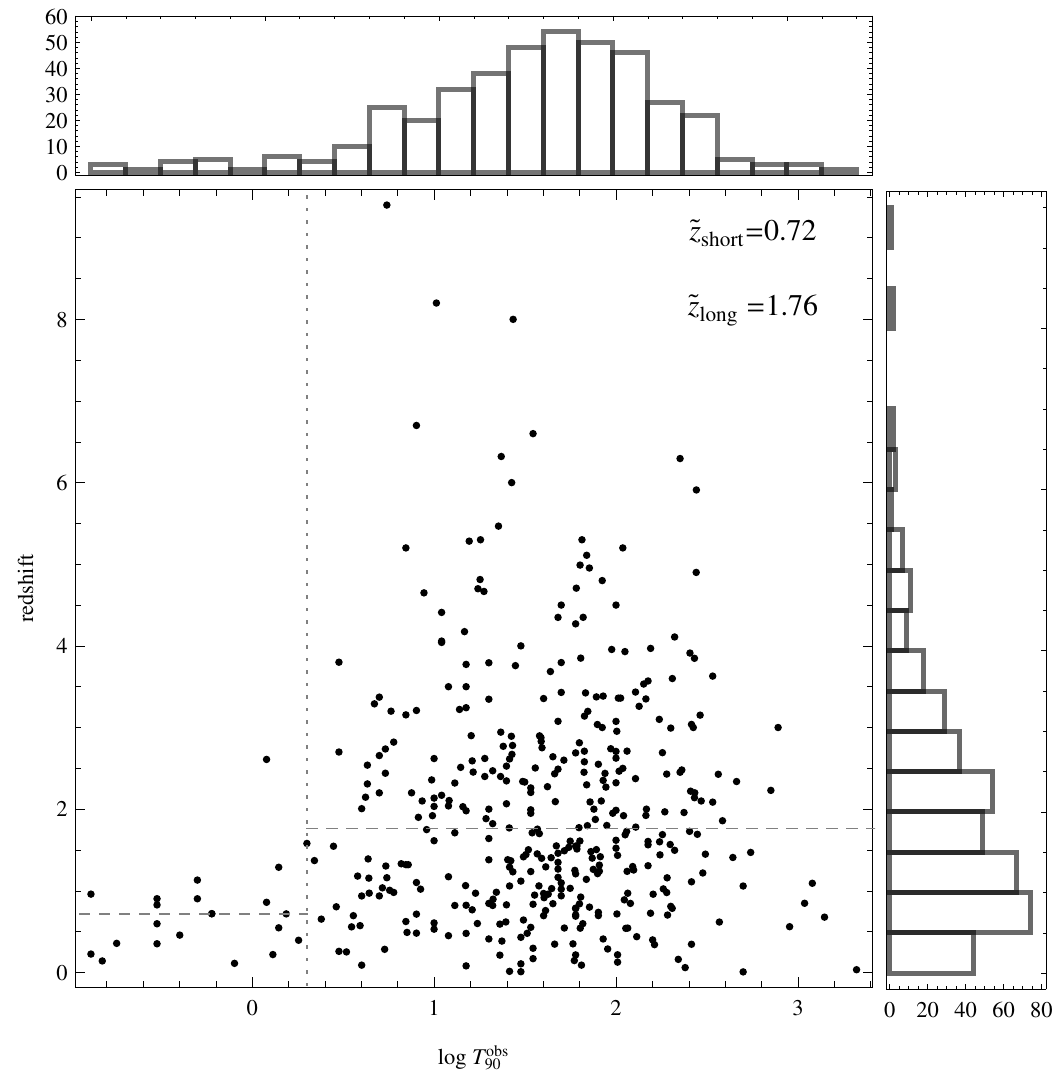}
\caption{A scatter plot of the redshifts versus the observed durations. Vertical dotted line marks the limitting value of $2\,{\rm s}$ between short and long GRBs, and the horizontal dashed lines denote the medians of the respective classes, with values written in the plot. All GRBs with known both $z$ and $T^{\rm obs}_{90}$ are shown.}
\label{fig1}
\end{figure}

\begin{figure}
\includegraphics[width=0.9\columnwidth]{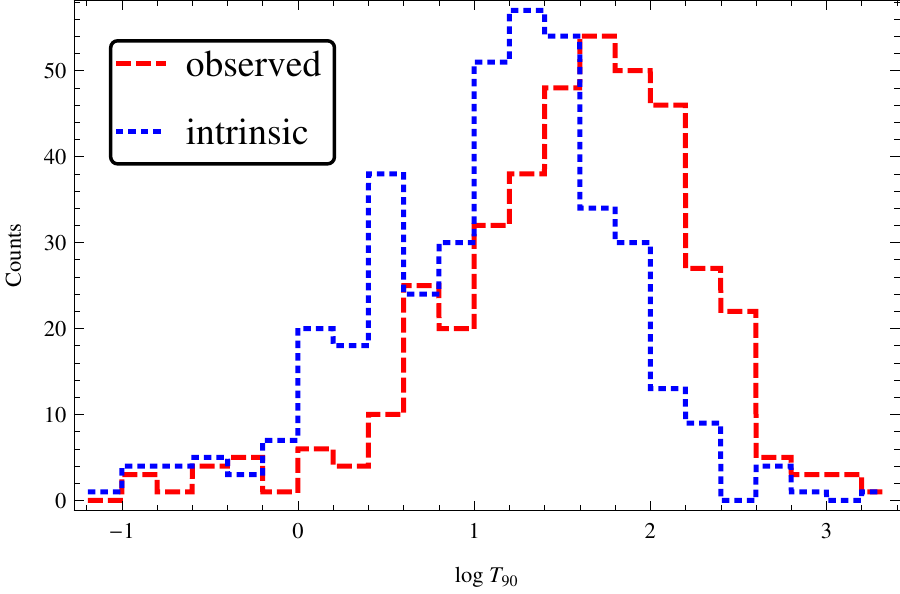}
\caption{Distributions of the observed (dashed red) and intrinsic (dotted blue) durations in the sample of all (408) GRBs.}
\label{fig2}
\end{figure}

A sample of 408 GRBs with measured both the observed durations $T^{\rm obs}_{90}$ and redshifts $z$ is used\footnote{\url{http://www.astro.caltech.edu/grbox/grbox.php}}. It contains 334 GRBs detected by {\it Swift}, constituting the second sample examined herein. The sample of all GRBs consists of 386 long GRBs and 22 short ones. The latter all come from {\it Swift} observations, except one that was detected by HETE (GRB040924). A scatter plot of the data on a redshift--logarithm of duration plane is drawn in Fig.~\ref{fig1}. The median redshifts for short and long GRBs are equal to $\tilde{z}_{\rm short}=0.72$ and $\tilde{z}_{\rm long}=1.76$, respectively. The intrinsic durations are calculated according to Eq.~(\ref{eq0}). Distributions of the $\log T_{90}$ for the observed and intrinsic durations are examined hereinafter, and are displayed in Fig.~\ref{fig2} for the sample of all GRBs.

\subsection{Fitting method}\label{fit}
Two standard fitting techniques are commonly applied: $\chi^2$ fitting and maximum likelihood method (ML). For the first, data needs to be binned, and despite various binning rules are known (e.g. Freedman-Diaconis, Scott, Knuth etc.), they still leave place for ambiguity, as it might happen that the fit may be statistically significant on a given significance level for a number of binnings \citep{huja2,koen,Tarnopolski}. The ML method is not affected by this issue and is therefore applied herein.  However, for display purposes, the binning was chosen based on the Freedman-Diaconis rule.

Having a distribution with a probability density function (PDF) given by $f=f(x;\theta)$ (possibly a mixture), where $\theta=\left\{\theta_i\right\}_{i=1}^p$ is a set of $p$ parameters, the log-likelihood function is defined as
\begin{equation}
\mathcal{L}_p(\theta)=\sum\limits_{i=1}^N\ln f(x_i;\theta),
\label{eq1}
\end{equation}
where $\left\{x_i\right\}_{i=1}^N$ are the datapoints from the sample to which a distribution is fitted. The fitting is performed by searching a set of parameters $\hat{\theta}$ for which the log-likelihood is maximized \citep{kendall}. When nested models are considered, the maximal value of the log-likelihood function, $\mathcal{L}_{\rm max}\equiv\mathcal{L}_p(\hat{\theta})$, increases when the number of parameters $p$ increases.

For nested as well as non-nested models, the Akaike information criterion ($AIC$) \citep{akaike,burnham,liddle,Tarnopolski2} may be applied. The $AIC$ is defined as
\begin{equation}
AIC=2p-2\mathcal{L}_{\rm max}.
\label{eq3}
\end{equation}
A preferred model is the one that minimizes $AIC$. The formulation of $AIC$ penalizes the use of an excessive number of parameters, hence discourages overfitting. It prefers models with fewer parameters, as long as the others do not provide a substantially better fit. The expression for $AIC$ consists of two competing terms: the first measuring the model complexity (number of free parameters) and the second measuring the goodness of fit (or more precisely, the lack of thereof). Among candidate models with $AIC_i$, let $AIC_{\rm min}$ denote the smallest. Then,
\begin{equation}
Pr_i=\exp\left(-\frac{\Delta_i}{2}\right),
\label{eq4}
\end{equation}
where $\Delta_i=AIC_i-AIC_{\rm min}$, can be interpreted as the relative (compared to $AIC_{\rm min}$) probability that the $i$-th model minimizes the $AIC$.\footnote{Relative probabilities of the models normalized to unity are called the Akaike weights, $w_i$. In Bayesian language, Akaike weight corresponds to the posterior probability of a model (under assumption of different prior probabilities; \citealt{biesiada}).}

The $AIC$ is suitable when $N/p$ is large, i.e. when $N/p>40$ \citep[][see also references therein]{burnham}. When this condition is not fulfilled, a second order bias correction is introduced, resulting in a small-sample version of the $AIC$, called $AIC_c$:
\begin{equation}
AIC_c=2p-2\mathcal{L}_{\rm max}+\frac{2p(p+1)}{N-p-1}.
\label{eq3a}
\end{equation}
The relative probability is computed similarly to when $AIC$ is used, i.e. Eq.~(\ref{eq4}) is valid when one takes $\Delta_i=AIC_{c,i}-AIC_{c,{\rm min}}$. Thence,
\begin{equation}
Pr_i=\exp\left(-\frac{AIC_{c,i}-AIC_{c,{\rm min}}}{2}\right).
\label{eq4a}
\end{equation}

It is important to note that this method allows to choose a model that is best among the chosen ones, but does not allow to state that this model is the best among all possible. Hence, the probabilities computed by means of Eq.~(\ref{eq4a}) are the relative, with respect to a model with $AIC_{c,{\rm min}}$, probabilities that the data is better described by a model with $AIC_{c,i}$. What is essential in assessing the goodness of a fit in the $AIC$ method is the difference, $\Delta_i=AIC_{c,i}-AIC_{c,{\rm min}}$, not the absolute values of the $AIC_{c,i}$.\footnote{The $AIC$ value contains scaling constants coming from the log-likelihood $\mathcal{L}$. One might conisder $\Delta_i=AIC_{c,i}-AIC_{c,{\rm min}}$ a rescaling transformation that forces the best model to have $\Delta_{\rm min}=0$, and so $\Delta_i$ are free of such scaling constants \citep{burnham}.} If $\Delta_i<2$, then there is substantial support for the $i$-th model, and the proposition that it is a more proper description is highly probable. If $2<\Delta_i<4$, then there is strong support for the $i$-th model. When $4<\Delta_i<7$, there is considerably less support, and models with $\Delta_i>10$ have essentially no support \citep{burnham,biesiada}.

\subsection{Distributions and their properties}\label{dist}

In nearly all researches conducted so far on the GRB duration distribution, three components were found to describe the observed distribution statistically better than a mixture of two components. However, in all previous analyses a mixture of standard (non-skewed) Gaussians was fitted. This might possibly lead to erroneous conclusions, as describing a non-symmetrical distribution by a mixture of symmetrical components will eventually lead to overfitting (some of the two-component skewed distributions considered below are characterized by fewer free parameters than a standard three-Gaussian). Moreover, \citet{zitouni} suggested that the duration distribution of long GRBs might not necessarily be symmetrical because of a non-symmetrical distribution of envelope masses of the progenitors. Since \citet{mcbreen} observed that the distribution of $\log T_{90}$ may be in form of a mixture of standard Gaussians, many authors followed this approach and also restrained the analysis to non-skewed normal distributions \citep{koshut,kouve2,horvath98,horvath02,horvath08,zhang2,horvath08,horvath09,huja,huja2,ripa,horvath10,koen,barnacka,Tarnopolski}. Therefore, in light of the suggestion of \citet{zitouni} that the $T_{90}$ distributions underlying the two well-established GRB classes \citep{kouve,woosley,nakar} may not be symmetrical \citep{Tarnopolski}, the following distributions are considered herein.

A mixture of $k$ standard normal (Gaussian) $\mathcal{N}(\mu,\sigma^2)$ distributions:
\begin{equation}
\begin{array}{l}
f^{(\mathcal{N})}_k(x) = \sum\limits_{i=1}^k A_i \varphi\left(\frac{x-\mu_i}{\sigma_i}\right) \\
\textcolor{white}{f^{(\mathcal{N})}_k(x)} = \sum\limits_{i=1}^k \frac{A_i}{\sqrt{2\pi}\sigma_i}\exp\left(-\frac{(x-\mu_i)^2}{2\sigma_i^2}\right),
\end{array}
\label{eq5}
\end{equation}
being described by $p=3k-1$ free parameters: $k$ pairs $(\mu_i,\sigma_i)$ and $k-1$ weights $A_i$, satysfying $\sum_{i=1}^k A_i=1$. Skewness of each component is $\gamma_1^{(\mathcal{N})}=0$.

A mixture of $k$ skew normal (SN) distributions \citep{ohagan,azzalini}:
\begin{equation}
\begin{array}{l}
f^{(SN)}_k(x) = \sum\limits_{i=1}^k 2A_i\varphi\left(\frac{x-\mu_i}{\sigma_i}\right)\Phi\left(\alpha_i\frac{x-\mu_i}{\sigma_i}\right) \\
\textcolor{white}{F^{(SN)}_k(x)} = \sum\limits_{i=1}^k \frac{2A_i}{\sqrt{2\pi}\sigma_i}\exp\left(-\frac{(x-\mu_i)^2}{2\sigma_i^2}\right) \times \\
\textcolor{white}{F^{(SN)}_k(x) = \sum\limits_{i=1}^k } \times \frac{1}{2}\left[1+\textrm{erf}\left(\alpha_i\frac{x-\mu_i}{\sqrt{2}\sigma_i}\right)\right],
\end{array}
\label{eq6}
\end{equation}
described by $p=4k-1$ parameters. Skewness of an SN distribution is
\begin{equation*}
\gamma_1^{(SN)}=\frac{4-\pi}{2}\frac{\left(\zeta\sqrt{2/\pi}\right)^3}{\left(1-2\zeta^2/\pi\right)^{3/2}},
\end{equation*}
where $\zeta=\frac{\alpha}{\sqrt{1+\alpha^2}}$, hence the skewness $\gamma_1^{(SN)}$ is solely based on the shape parameter $\alpha$, and is limited to the interval $\left(-1,1\right)$. The mean is given by $\mu+\sigma\zeta\sqrt{\frac{2}{\pi}}$. When $\alpha=0$, the SN distribution is reduced to a standard Gaussian, $\mathcal{N}(\mu,\sigma^2)$, due to $\Phi(0)=1/2$.

A mixture of $k$ sinh-arcsinh (SAS) distributions \citep{jones}:
\begin{equation}
\begin{array}{l}
f^{(SAS)}_k(x) = \sum\limits_{i=1}^k \frac{A_i}{\sigma_i}\left[1+\left(\frac{x-\mu_i}{\sigma_i}\right)^2\right]^{-\frac{1}{2}} \times \\
\textcolor{white}{F^{(SAS)}_k(x)} \times \beta_i \cosh\left[\beta_i\sinh^{-1}\left(\frac{x-\mu_i}{\sigma_i}\right)-\delta_i\right]\times \\
\textcolor{white}{F^{(SAS)}_k(x)} \times\exp\left[-\frac{1}{2}\sinh\left[\beta_i\sinh^{-1}\left(\frac{x-\mu_i}{\sigma_i}\right)-\delta_i\right]^2\right],
\end{array}
\label{eq7}
\end{equation}
being described by $p=5k-1$ parameters. It turns out that skewness of the SAS distribution increases with increasing $\delta$, positive skewness corresponding to $\delta>0$. Tailweight decreases with increasing $\beta$, $\beta<1$ yielding heavier tails than the normal distribution, and $\beta>1$ yielding lighter tails. With $\delta=0$ and $\beta=1$, the SAS distribution reduces to a standard Gaussian, $\mathcal{N}(\mu,\sigma^2)$. Skewness of a SAS distribution is
\begin{equation*}
\gamma_1^{(SAS)}=\frac{1}{4}\left[ \sinh\left(\frac{3\delta}{\beta}\right)P_{3/\beta} - 3\sinh\left(\frac{\delta}{\beta}\right)P_{1/\beta} \right],
\end{equation*}
where
\begin{equation*}
P_q=\frac{e^{1/4}}{\sqrt{8\pi}}\left[ K_{(q+1)/2}(1/4) + K_{(q-1)/2}(1/4) \right].
\end{equation*}
Here, $K$ is the modified Bessel function of the second kind. The mean is given by $\mu+\sigma\sinh(\delta/\beta)P_{1/\beta}$.

A mixture of $k$ alpha-skew-normal (ASN) distributions \citep{elal}:
\begin{equation}
\begin{array}{l}
f^{(ASN)}_k(x)=\sum\limits_{i=1}^k A_i\frac{\left(1-\alpha_i\frac{x-\mu_i}{\sigma_i}\right)^2+1}{2+\alpha_i^2} \varphi\left(\frac{x-\mu_i}{\sigma_i}\right) \\
\textcolor{white}{F^{(ASN)}_k(x)} = \sum\limits_{i=1}^k A_i\frac{\left(1-\alpha_i\frac{x-\mu_i}{\sigma_i}\right)^2+1}{2+\alpha_i^2}\times \\
\textcolor{white}{F^{(ASN)}_k(x) = \sum\limits_{i=1}^k} \times\frac{1}{\sqrt{2\pi}\sigma_i}\exp\left(-\frac{(x-\mu_i)^2}{2\sigma_i^2}\right),
\end{array}
\label{eq8}
\end{equation}
described by $p=4k-1$ parameters. Skewness of an ASN distribution is
\begin{equation*}
\gamma_1^{(ASN)}=\frac{12\alpha^5+8\alpha^3}{(3\alpha^4+4\alpha^2+4)^{3/2}},
\end{equation*}
and is limited to the interval $(-0.811,0.811)$. The mean is given by $\mu-\frac{2\alpha\sigma}{2+\alpha^2}$. For $\alpha\in(-1.34,1.34)$ the distribution is unimodal, and bimodal otherwise.

\section{Study of the complete $z$ sample}\label{res1}
The biggest number of free parameters among the examined PDFs is $p=14$ in the case of a 3-SAS. Combined with $N=408$ for all GRBs, or $N=334$ for the {\it Swift} subsample, this implies $N/p<40$. Therefore, in what follows the $AIC_c$ is used instead of the $AIC$.

First, the sample of 408 GRBs with measured both redshift and duration is examined. The PDFs, given by Eq.~(\ref{eq5})--(\ref{eq8}), with $k=2$ or 3, are fitted to the $\log T_{90}$ distributions using the ML method from Sect.~\ref{fit}. Next, the $AIC_c$ is calculated according to Eq.~(\ref{eq3a}). The best fit among the examined is the one that yields the smallest $AIC_c$.

\begin{figure}
\includegraphics[width=\columnwidth]{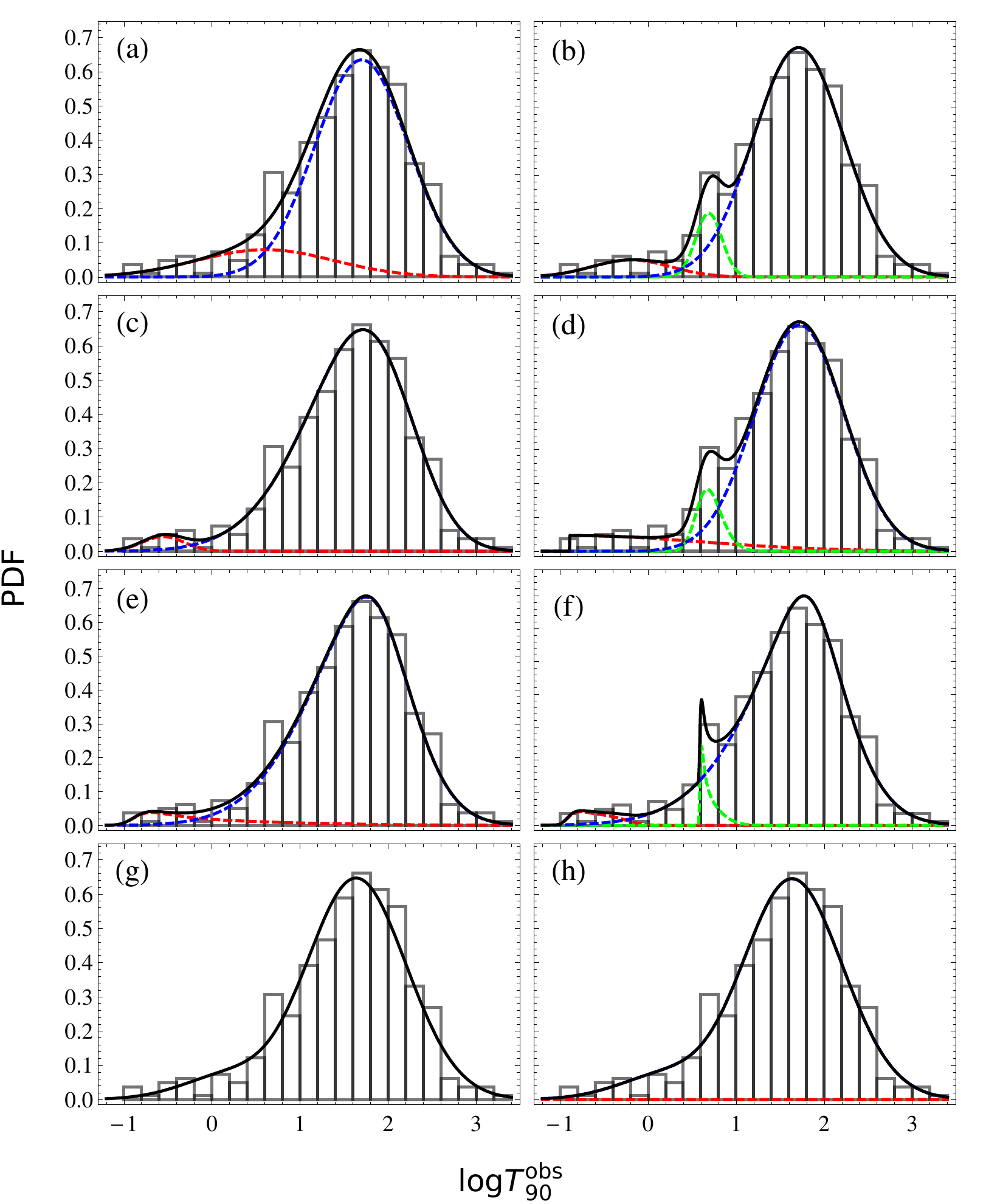}
\caption{Distributions fitted to $\log T^{\rm obs}_{90}$ data of all GRBs. Color dashed curves are the components of the (black solid) mixture distribution. The panels show a mixture of (a) two standard Gaussians (2-G), (b) three standard Gaussians (3-G), (c) two skew-normal (2-SN), (d) three skew-normal (3-SN), (e) two sinh-arcsinh (2-SAS), (f) three sinh-arcsinh (3-SAS), (g) one alpha-skew-normal (1-ASN), and (h) two alpha-skew-normal (2-ASN) distributions.}
\label{fig3}
\end{figure}

\begin{figure}
\includegraphics[width=\columnwidth]{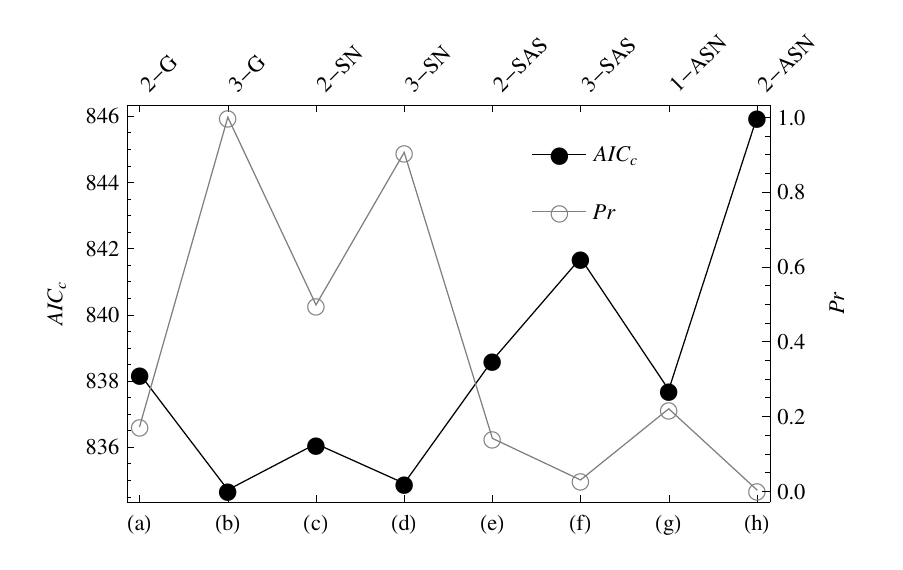}
\caption{$AIC_c$ and relative probability, $Pr$, for the models examined and for observed durations.}
\label{fig4}
\end{figure}

The results of the fitting procedure applied to the observed durations are gathered in Table~\ref{tbl1}, and the fits, in graphical form, are displayed in Fig.~\ref{fig3}. Contrary to previous reasearches, all of the three-component PDFs (3-G, 3-SN, and 3-SAS, where the support for the latter is weak) are trimodal, and the third peak is located in the area of the presumed intermediate GRB class, i.e. within the range $2-10\,{\rm s}$. To assess its significance more easily, the $AIC_c$ and relative probabilities are plotted in Fig.~\ref{fig4}. The PDF with minimal $AIC_c$ is a conventional 3-G, and the second best fit is a 3-SN, with a relative probability of 90.6\%. A 2-SN, however, has substantial support, too, due to $\Delta_{\rm 2-SN}=1.393$. The remaining two-component fits (2-G and 2-SAS), as well as a 1-ASN, yield a strong support having $2<\Delta_i<4$, but the evidence is weaker than for the former three models. The remaining, 3-SAS and 2-ASN, have considerably less or no support.

\begin{figure}
\includegraphics[width=\columnwidth]{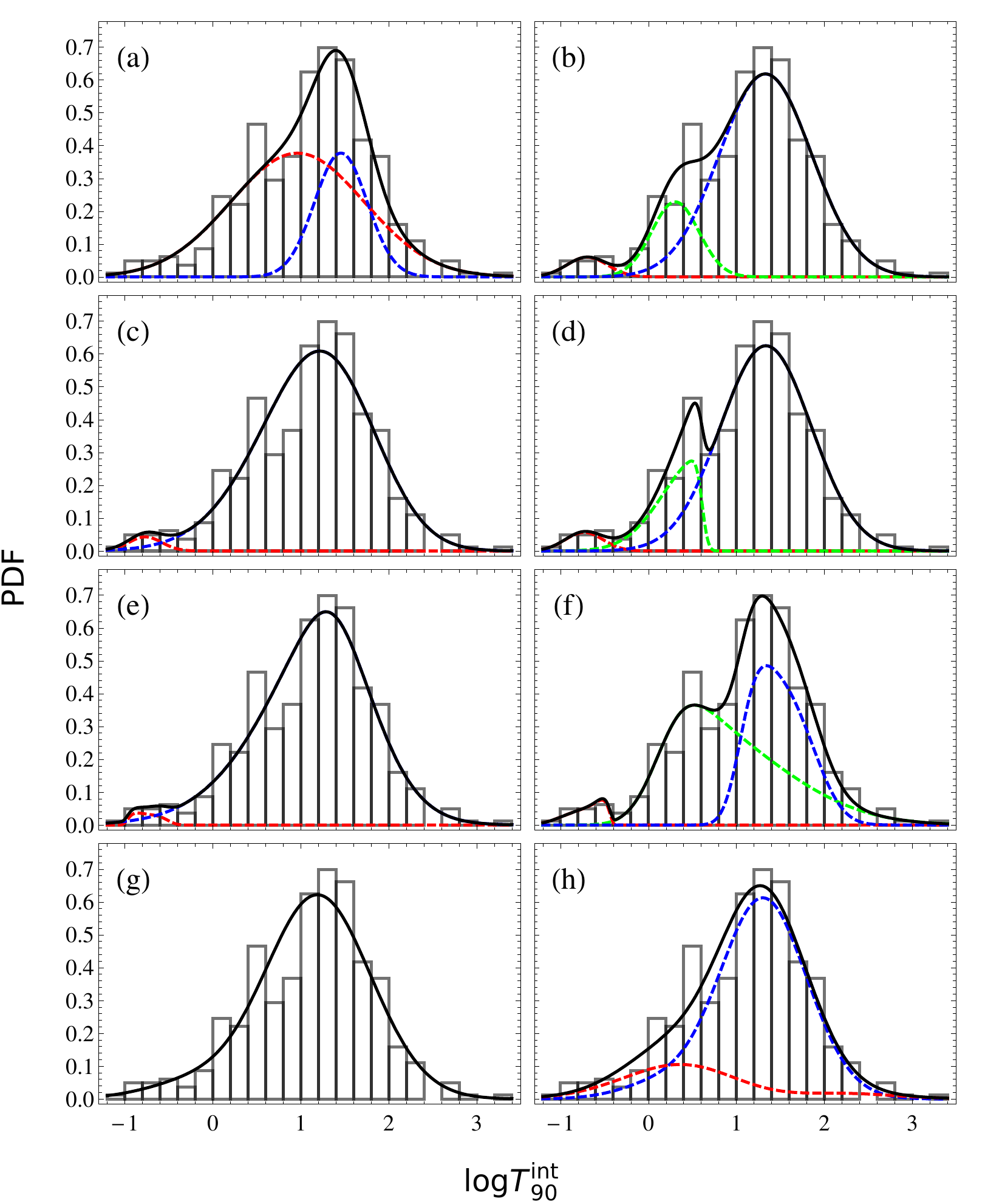}
\caption{The same as Fig.~\ref{fig3}, but for $\log T^{\rm int}_{90}$.}
\label{fig5}
\end{figure}

\begin{figure}
\includegraphics[width=\columnwidth]{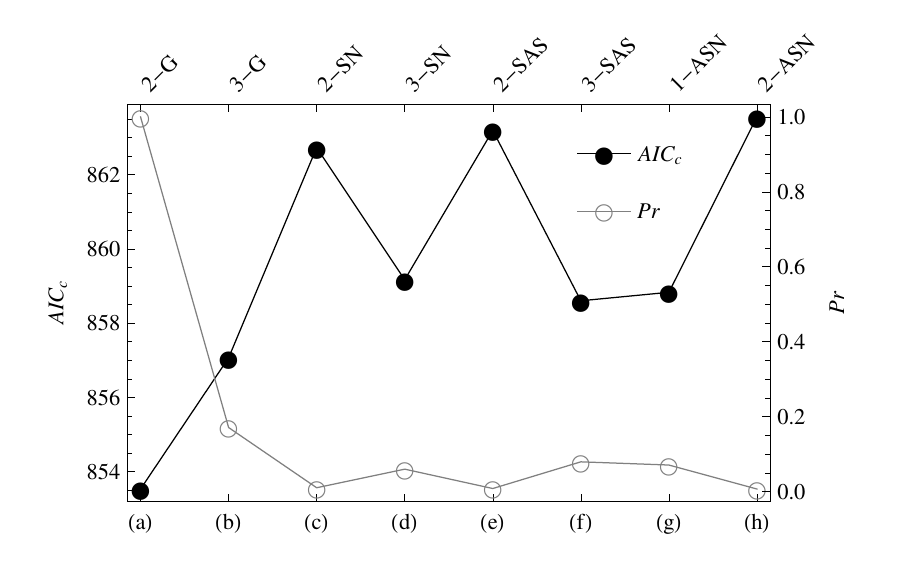}
\caption{The same as Fig.~\ref{fig4}, but for intrinsic durations.}
\label{fig6}
\end{figure}

The picture revealed by the rest frame duration distribution, $T^{\rm int}_{90}$, is different. As displayed in Fig.~\ref{fig5}, the 3-SN and 3-SAS are also trimodal, and the 3-G, with the durations being systematically shifted left-wards comparing with the observed durations, lost its third peak, leaving a bimodal distribution with a prominent shoulder in the area of the presumed intermediate GRBs. The parameters of the fits are gathered in Table~\ref{tbl2}. A remarkably different picture, compared to the result of the $T^{\rm obs}_{90}$ analysis, follows from the $AIC_c$ plot in Fig.~\ref{fig6}. It turns out that the intrinsic durations are best described by a conventional 2-G; the second best model is a 3-G, having a relative probability of 17.1\%. The other models have considerably less or almost no support. This suggests that the intrinsic $T_{90}$ distribution may be indeed bimodal.

To conclude this Section, the $T_{90}^{\rm obs}$ distribution is possibly trimodal, and in the rest frame, due to the properties of Eq.~(\ref{eq1}), it turns into a bimodal.

\begin{table*}
\small
\caption{Parameters of the fits for the observed durations. Label corresponds to labels from Fig.~\ref{fig3} and \ref{fig4}. The smallest $AIC_c$ is marked in bold, and $p$ is the number of parameters in a model.}
\label{tbl1}
\centering
\begin{tabular}{@{}cccccccccccccc@{}}
\hline\hline
  Label & Dist. & $i$ & $\mu_i$ & $\sigma_i$ & $\alpha_i$ & $\delta_i$ & $\beta_i$ & $A_i$ & $\mathcal{L}_{\rm max}$ & $AIC_c$ & $\Delta AIC_c$ & $Pr$ & $p$ \\
  \hline
\multirow{2}{*}{(a)} & \multirow{2}{*}{2-G} & 1 & 0.607 & 0.777 & --- & --- & --- & 0.156 & \multirow{2}{*}{$-414.032$} & \multirow{2}{*}{838.214} & \multirow{2}{*}{3.505} & \multirow{2}{*}{0.173} & \multirow{2}{*}{5} \\
    &     & 2 &  1.702 & 0.531 & --- & --- & --- & 0.844 & & & & & \\
  \hline
    &     & 1 & $-0.192$ & 0.447 & --- & --- & --- & 0.057 & & & & & \\
(b) & 3-G & 2 &  0.690   & 0.149 & --- & --- & --- & 0.070 & $-409.174$ & \textbf{834.709} & 0 & 1 & 8 \\
    &     & 3 &  1.710   & 0.515 & --- & --- & --- & 0.873 & & & & & \\
  \hline
\multirow{2}{*}{(c)} & \multirow{2}{*}{2-SN} & 1 & $-0.557$ & 0.240 & 0.013 & --- & --- & 0.026 & \multirow{2}{*}{$-410.911$} & \multirow{2}{*}{836.102} & \multirow{2}{*}{1.393} & \multirow{2}{*}{0.498} & \multirow{2}{*}{7} \\
    &      & 2 &  2.178 & 0.857 & $-1.732$ & --- & --- & 0.974 & & & & & \\
  \hline
    &      & 1 & $-0.890$ & 1.471 &  $1.025\times 10^3$ & --- & --- & 0.085 & & & & & \\
(d) & 3-SN & 2 &  0.564   & 0.205 &  1.701    & --- & --- & 0.066 & $-406.120$ & 834.906 & 0.197 & 0.906 & 11 \\
    &      & 3 &  1.688   & 0.508 &  0.078    & --- & --- & 0.849 & & & & & \\
  \hline
\multirow{2}{*}{(e)} & \multirow{2}{*}{2-SAS} & 1 & $-0.868$ & 0.350 & --- & 1.261 & 0.919 & 0.046 & \multirow{2}{*}{$-410.085$} & \multirow{2}{*}{838.622} & \multirow{2}{*}{3.913} & \multirow{2}{*}{0.141} & \multirow{2}{*}{9} \\
    &       & 2 & 1.753 & 0.516 & --- & $-0.185$ & 0.914 & 0.954 & & & & & \\
  \hline
    &       & 1 & $-1.026$ & 0.428 & --- &  1.893    & 2.224 & 0.024 & & & & & \\
(f) & 3-SAS & 2 & 0.574    & 0.028 & --- &  1.866    & 0.960 & 0.033 & $-406.322$ & 841.712 & 7.003 & 0.030 & 14 \\
    &       & 3 & 1.773    & 0.451 & --- &  $-0.192$ & 0.838 & 0.943 & & & & & \\
  \hline
(g) & 1-ASN & 1 & 1.099 & 0.634 & $-1.017$ & --- & --- & 1 & $-415.837$ & 837.733 & 3.024 & 0.220 & 3 \\
  \hline
\multirow{2}{*}{(h)} & \multirow{2}{*}{2-ASN} & 1 & 1.096 & 0.636 & $-1.017$ & --- & --- & 0.999 & \multirow{2}{*}{$-415.844$} & \multirow{2}{*}{845.969} & \multirow{2}{*}{11.260} & \multirow{2}{*}{0.004} & \multirow{2}{*}{7} \\
    &       & 2 & 0.237 & 1.313 & $-0.946$ & --- & --- & 0.001 & & & & & \\
  \hline
\end{tabular}
\end{table*}

\begin{table*}
\small
\caption{Parameters of the fits for the intrinsic durations. Label corresponds to labels from Fig.~\ref{fig5} and \ref{fig6}. The smallest $AIC_c$ is marked in bold, and $p$ is the number of parameters in a model.}
\label{tbl2}
\centering
\begin{tabular}{@{}cccccccccccccc@{}}
\hline\hline
  Label & Dist. & $i$ & $\mu_i$ & $\sigma_i$ & $\alpha_i$ & $\delta_i$ & $\beta_i$ & $A_i$ & $\mathcal{L}_{\rm max}$ & $AIC_c$ & $\Delta AIC_c$ & $Pr$ & $p$ \\
  \hline
\multirow{2}{*}{(a)} & \multirow{2}{*}{2-G} & 1 & 0.962 & 0.759 & --- & --- & --- & 0.717 & \multirow{2}{*}{$-421.691$} & \multirow{2}{*}{\textbf{853.532}} & \multirow{2}{*}{0} & \multirow{2}{*}{1} & \multirow{2}{*}{5} \\
    &     & 2 &  1.456 & 0.299 & --- & --- & --- & 0.283 & & & & & \\
  \hline
    &     & 1 & $-0.696$ & 0.232 & --- & --- & --- & 0.035 & & & & & \\
(b) & 3-G & 2 &  0.305   & 0.268 & --- & --- & --- & 0.154 & $-420.353$ & 857.066 & 3.534 & 0.171 & 8 \\
    &     & 3 &  1.330   & 0.524 & --- & --- & --- & 0.811 & & & & & \\
  \hline
\multirow{2}{*}{(c)} & \multirow{2}{*}{2-SN} & 1 & $-0.769$ & 0.188 & 0.015 & --- & --- & 0.020 & \multirow{2}{*}{$-424.224$} & \multirow{2}{*}{862.728} & \multirow{2}{*}{9.196} & \multirow{2}{*}{0.010} & \multirow{2}{*}{7} \\
    &      & 2 &  1.662 & 0.834 & $-1.259$ & --- & --- & 0.980 & & & & & \\
  \hline
    &      & 1 & $-0.735$ & 0.216 &  0.025 & --- & --- & 0.031 & & & & & \\
(d) & 3-SN & 2 &  0.610   & 0.439 &  0.439 & --- & --- & 0.158 & $-418.252$ & 859.170 & 5.638 & 0.060 & 11 \\
    &      & 3 &  1.333   & 0.518 &  0.006 & --- & --- & 0.811 & & & & & \\
  \hline
\multirow{2}{*}{(e)} & \multirow{2}{*}{2-SAS} & 1 & $-1.527$ & 0.288 & --- & 7.959 & 4.655 & 0.015 & \multirow{2}{*}{$-422.369$} & \multirow{2}{*}{863.190} & \multirow{2}{*}{9.658} & \multirow{2}{*}{0.008} & \multirow{2}{*}{9} \\
    &       & 2 & 1.289 & 0.527 & --- & $-0.172$ & 0.871 & 0.985 & & & & & \\
  \hline
    &       & 1 & $-0.298$ & 0.168 & --- &  $-2.774$ & 1.790 & 0.033 & & & & & \\
(f) & 3-SAS & 2 & 0.453    & 0.630 & --- &  0.586    & 1.056 & 0.556 & $-414.767$ & 858.603 & 5.071 & 0.079 & 14 \\
    &       & 3 & 1.322    & 0.444 & --- &  0.376    & 1.317 & 0.411 & & & & & \\
  \hline
(g) & 1-ASN & 1 & 0.699 & 0.654 & $-0.837$ & --- & --- & 1 & $-426.386$ & 858.831 & 5.299 & 0.071 & 3 \\
  \hline
\multirow{2}{*}{(h)} & \multirow{2}{*}{2-ASN} & 1 & 0.900 & 0.538 & $-0.829$ & --- & --- & 0.810 & \multirow{2}{*}{$-424.638$} & \multirow{2}{*}{863.556} & \multirow{2}{*}{10.024} & \multirow{2}{*}{0.007} & \multirow{2}{*}{7} \\
    &       & 2 & 1.090 & 0.747 & 1.338 & --- & --- & 0.190 & & & & & \\
  \hline
\end{tabular}
\end{table*}

\section{Study of the \textit{Swift} subsample}\label{res2}
The reason for examinig separately a smaller sample of 334 GRBs detected only by {\it Swift} is the fact that the $T_{90}$ distributions and other features, e.g. sensitivity in different energy bands, are detector dependent \citep[e.g.,][]{horvath06,horvath09,huja,ripa,horvath10,bromberg2,zitouni,Tarnopolski2,Tarnopolski3}, and thus the sample examined in the previous Section~\ref{res1} might be biased. The majority of GRBs with known redshift comes from {\it Swift}, and hence one might consider the detections made by other satellites a contamination that falsifies the outcome. Therefore, it is desired to investigate a sample in which all observations were made by the same instrument.

\begin{figure}
\includegraphics[width=\columnwidth]{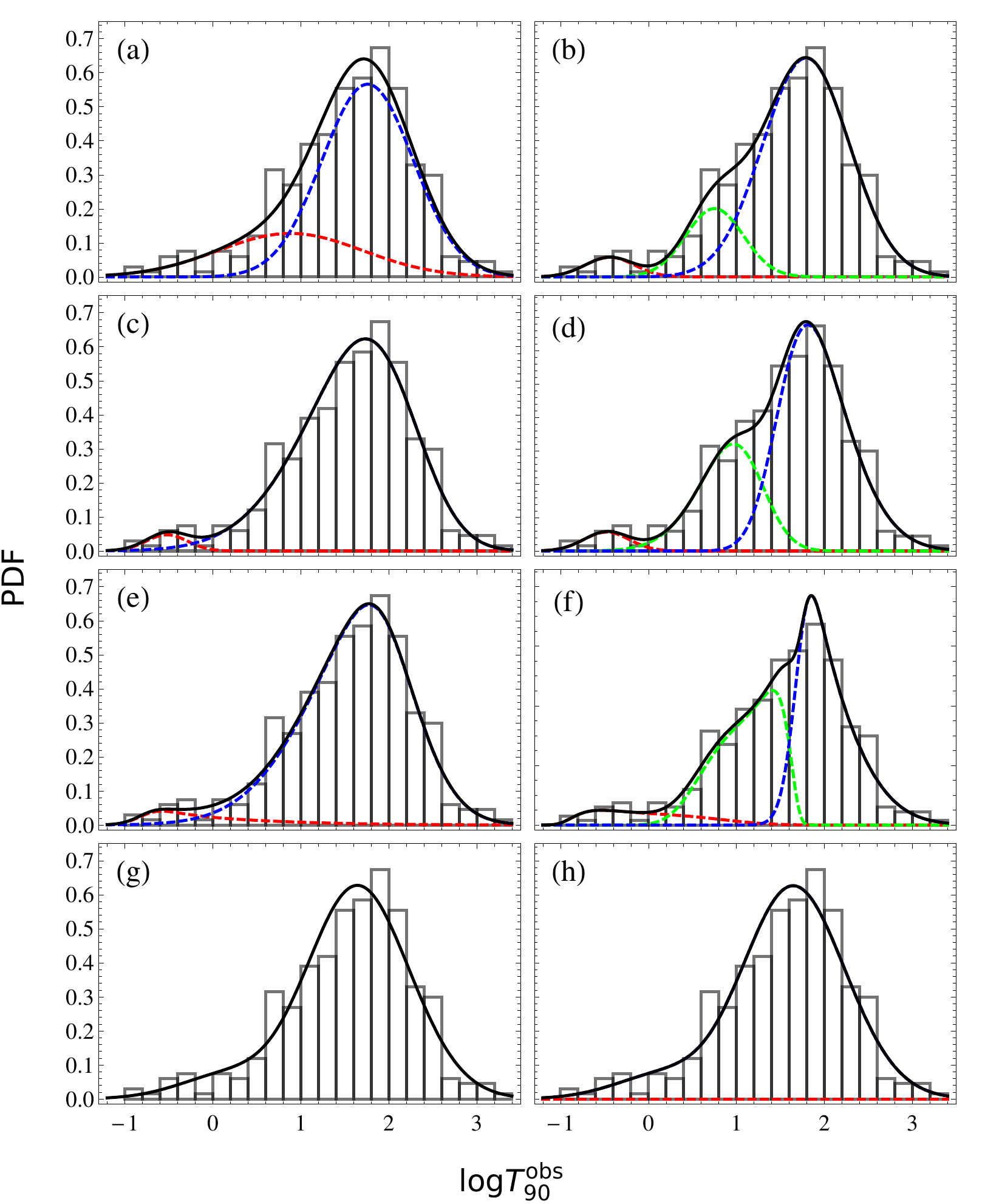}
\caption{The same as Fig.~\ref{fig3}, but for $\log T^{\rm obs}_{90}$ of the {\it Swift} subsample.}
\label{fig7}
\end{figure}

The analysis is performed in the same way as it was done in Sect.~\ref{res1}. Again, the observed durations are examined first. A striking difference is that all of the fits are at most bimodal, and unimodal when a 2-G, 1-ASN or 2-ASN is considered. The parameters of the fits are gathered in Table~\ref{tbl3}, and the fitted curves are displayed in Fig.~\ref{fig7}. The uni- or bimodality is consistent with previous analyses performed on more complete samples of {\it Swift} GRBs \citep{horvath08,huja,zitouni,Tarnopolski2}, and the curves for three-component fits show a prominent shoulder on the left-hand side of the peak related to long GRBs.

The $AIC$ indicates that the distribution of $\log T_{90}^{\rm obs}$ is best described by a 3-G. The next two best fits, a 1-ASN and a 2-SN, have a $\Delta_i<2$, and hence yield strong support in their favor. Next, a 2-G has a relative probability of 24.3\% of being a more proper model. The remaining models have considerably less support. It follows that in case of the observed durations one cannot discern reliably the best description among a one- or two-component PDFs, what is also consistent with the previous analyses, as the {\it Swift} detection rate is heavily biased towards long GRBs (the ratio of short to long GRBs is $<1:14$), hence the sample is strongly dominated by long GRBs. Hence, combined with the relatively low number of redshift-equipped GRBs, it appears that due to this domination is unambiguous, in terms of modality, classification of the $T_{90}^{\rm int}$ distribution is uncertain.

When the intrinsic durations are considered, there appear some trimodal fits (see Fig.~\ref{fig9}). Surprisingly, the model with the lowest $AIC$ is a bimodal 2-ASN, while the second best fit is achieved by a (also bimodal) 3-G, having a relative probability of 13.9\%. The remaining models have signinificantly less support (compare with Table~\ref{tbl4} and Fig.~\ref{fig10}). The 2-ASN consists of a bimodal and a unimodal component. The former consists of two peaks with comparable height, and is visually very symmetric. The latter is skewed, with its mode placed near the peak of the bimodal component that corresponds to long GRBs. Hence, the overall role of the unimodal component is to rescale the bimodal one in a nonlinear way in order to follow the data. The structure of this fit is unusual and unexpected, as in the previous samples the 2-ASN model did not perform very well, being one of the worst fits.

\begin{table*}
\small
\caption{Parameters of the fits for the observed {\it Swift} durations. Label corresponds to labels from Fig.~\ref{fig7} and \ref{fig8}. The smallest $AIC_c$ is marked in bold, and $p$ is the number of parameters in a model.}
\label{tbl3}
\centering
\begin{tabular}{@{}cccccccccccccc@{}}
\hline\hline
  Label & Dist. & $i$ & $\mu_i$ & $\sigma_i$ & $\alpha_i$ & $\delta_i$ & $\beta_i$ & $A_i$ & $\mathcal{L}_{\rm max}$ & $AIC_c$ & $\Delta AIC_c$ & $Pr$ & $p$ \\
  \hline
\multirow{2}{*}{(a)} & \multirow{2}{*}{2-G} & 1 & 0.883 & 0.822 & --- & --- & --- & 0.263 & \multirow{2}{*}{$-351.758$} & \multirow{2}{*}{713.698} & \multirow{2}{*}{2.826} & \multirow{2}{*}{0.243} & \multirow{2}{*}{5} \\
    &     & 2 &  1.760 & 0.519 & --- & --- & --- & 0.737 & & & & & \\
  \hline
    &     & 1 & $-0.439$ & 0.280 & --- & --- & --- & 0.041 & & & & & \\
(b) & 3-G & 2 &  0.751   & 0.328 & --- & --- & --- & 0.165 & $-347.214$ & \textbf{710.872} & 0 & 1 & 8 \\
    &     & 3 &  1.792   & 0.493 & --- & --- & --- & 0.794 & & & & & \\
  \hline
\multirow{2}{*}{(c)} & \multirow{2}{*}{2-SN} & 1 & $-0.499$ & 0.236 & $-0.125$ & --- & --- & 0.028 & \multirow{2}{*}{$-348.914$} & \multirow{2}{*}{712.172} & \multirow{2}{*}{1.300} & \multirow{2}{*}{0.521} & \multirow{2}{*}{7} \\
    &      & 2 &  2.218 & 0.901 & $-1.818$ & --- & --- & 0.972 & & & & & \\
  \hline
    &      & 1 & $-0.507$ & 0.261 &  0.152    & --- & --- & 0.037 & & & & & \\
(d) & 3-SN & 2 &  1.245   & 0.515 &  $-1.648$ & --- & --- & 0.295 & $-346.393$ & 715.606 & 4.734 & 0.094 & 11 \\
    &      & 3 &  1.500   & 0.587 &  2.028    & --- & --- & 0.669 & & & & & \\
  \hline
\multirow{2}{*}{(e)} & \multirow{2}{*}{2-SAS} & 1 & $-0.695$ & 0.409 & --- & 0.921 & 0.897 & 0.051 & \multirow{2}{*}{$-348.580$} & \multirow{2}{*}{715.715} & \multirow{2}{*}{4.843} & \multirow{2}{*}{0.089} & \multirow{2}{*}{9} \\
    &       & 2 & 1.780 & 0.540 & --- & $-0.201$ & 0.923 & 0.949 & & & & & \\
  \hline
    &       & 1 & $-1.645$ & 0.776 & --- &  3.578    & 2.502 & 0.071 & & & & & \\
(f) & 3-SAS & 2 & 1.920    & 0.543 & --- &  $-2.538$ & 2.232 & 0.397 & $-345.483$ & 720.283 & 9.411 & 0.009 & 14 \\
    &       & 3 & 1.831    & 0.245 & --- &  0.526    & 0.900 & 0.532 & & & & & \\
  \hline
(g) & 1-ASN & 1 & 1.088 & 0.654 & $-1.022$ & --- & --- & 1 & $-352.883$ & 711.845 & 0.973 & 0.614 & 3 \\
  \hline
\multirow{2}{*}{(h)} & \multirow{2}{*}{2-ASN} & 1 & 1.086 & 0.655 & $-1.026$ & --- & --- & 0.999 & \multirow{2}{*}{$-352.888$} & \multirow{2}{*}{720.121} & \multirow{2}{*}{9.249} & \multirow{2}{*}{0.010} & \multirow{2}{*}{7} \\
    &       & 2 & 0.546 & 1.210 & $-0.736$ & --- & --- & 0.001 & & & & & \\
  \hline
\end{tabular}
\end{table*}

\begin{figure}
\includegraphics[width=\columnwidth]{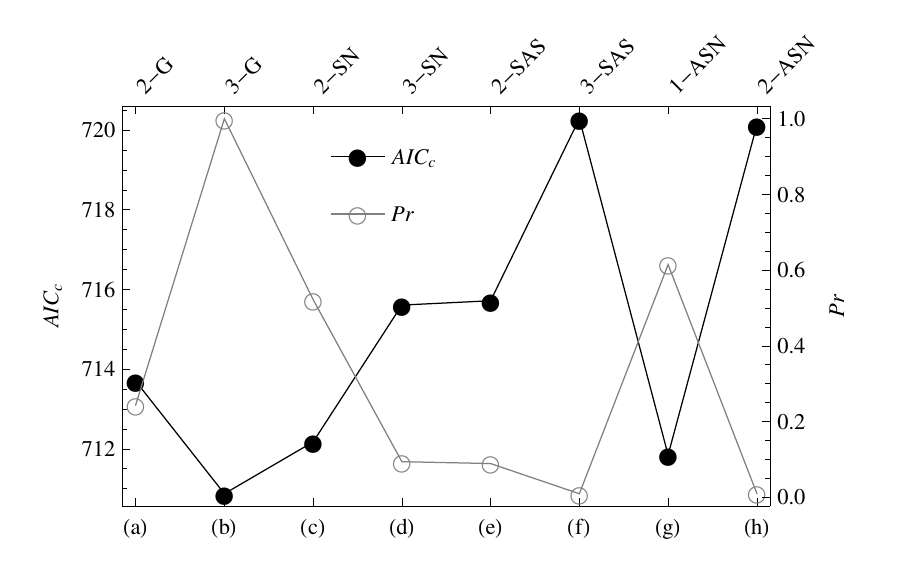}
\caption{The same as Fig.~\ref{fig4}, but for observed durations and {\it Swift} GRBs.}
\label{fig8}
\end{figure}

\begin{table*}
\small
\caption{Parameters of the fits for the intrinsic {\it Swift} durations. Label corresponds to labels from Fig.~\ref{fig9} and \ref{fig10}. The smallest $AIC_c$ is marked in bold, and $p$ is the number of parameters in a model.}
\label{tbl4}
\centering
\begin{tabular}{@{}cccccccccccccc@{}}
\hline\hline
  Label & Dist. & $i$ & $\mu_i$ & $\sigma_i$ & $\alpha_i$ & $\delta_i$ & $\beta_i$ & $A_i$ & $\mathcal{L}_{\rm max}$ & $AIC_c$ & $\Delta AIC_c$ & $Pr$ & $p$ \\
  \hline
\multirow{2}{*}{(a)} & \multirow{2}{*}{2-G} & 1 & 0.105 & 0.605 & --- & --- & --- & 0.143 & \multirow{2}{*}{$-360.409$} & \multirow{2}{*}{731.001} & \multirow{2}{*}{7.822} & \multirow{2}{*}{0.020} & \multirow{2}{*}{5} \\
    &     & 2 &  1.251 & 0.598 & --- & --- & --- & 0.857 & & & & & \\
  \hline
    &     & 1 & $-0.757$ & 0.193 & --- & --- & --- & 0.020 & & & & & \\
(b) & 3-G & 2 &  1.035   & 0.715 & --- & --- & --- & 0.788 & $-355.344$ & 727.130 & 3.951 & 0.139 & 8 \\
    &     & 3 &  1.493   & 0.275 & --- & --- & --- & 0.192 & & & & & \\
  \hline
\multirow{2}{*}{(c)} & \multirow{2}{*}{2-SN} & 1 & $-0.505$ & 0.299 & $-3.146$ & --- & --- & 0.026 & \multirow{2}{*}{$-358.784$} & \multirow{2}{*}{731.912} & \multirow{2}{*}{8.733} & \multirow{2}{*}{0.013} & \multirow{2}{*}{7} \\
    &      & 2 &  1.569 & 0.794 & $-0.938$ & --- & --- & 0.974 & & & & & \\
  \hline
    &      & 1 & $-0.496$ & 0.317 &  $-4.498$ & --- & --- & 0.022 & & & & & \\
(d) & 3-SN & 2 &  1.033   & 0.709 &  0.709    & --- & --- & 0.793 & $-355.108$ & 733.036 & 9.857 & 0.007 & 11 \\
    &      & 3 &  1.483   & 0.273 &  0.051    & --- & --- & 0.185 & & & & & \\
  \hline
\multirow{2}{*}{(e)} & \multirow{2}{*}{2-SAS} & 1 & $-0.795$ & 0.467 & --- & 0.708 & 1.086 & 0.032 & \multirow{2}{*}{$-358.638$} & \multirow{2}{*}{735.831} & \multirow{2}{*}{12.652} & \multirow{2}{*}{0.002} & \multirow{2}{*}{9} \\
    &       & 2 & 1.267 & 0.563 & --- & $-0.141$ & 0.896 & 0.968 & & & & & \\
  \hline
    &       & 1 & $-0.354$ & 0.056 & --- &  $-3.073$ & 1.322 & 0.038 & & & & & \\
(f) & 3-SAS & 2 & 0.408    & 0.686 & --- &  0.664    & 1.123 & 0.592 & $-350.300$ & 729.917 & 6.738 & 0.034 & 14 \\
    &       & 3 & 1.230    & 0.632 & --- &  0.664    & 1.843 & 0.370 & & & & & \\
  \hline
(g) & 1-ASN & 1 & 0.679 & 0.680 & $-0.801$ & --- & --- & 1 & $-360.796$ & 727.665 & 4.486 & 0.106 & 3 \\
  \hline
\multirow{2}{*}{(h)} & \multirow{2}{*}{2-ASN} & 1 & 0.879 & 0.324 & $-4.251\times 10^4$ & --- & --- & 0.270 & \multirow{2}{*}{$-354.418$} & \multirow{2}{*}{\textbf{723.179}} & \multirow{2}{*}{0} & \multirow{2}{*}{1} & \multirow{2}{*}{7} \\
    &       & 2 & 0.706 & 0.661 & $-1.148$ & --- & --- & 0.730 & & & & & \\
  \hline
\end{tabular}
\end{table*}

\begin{figure}
\includegraphics[width=\columnwidth]{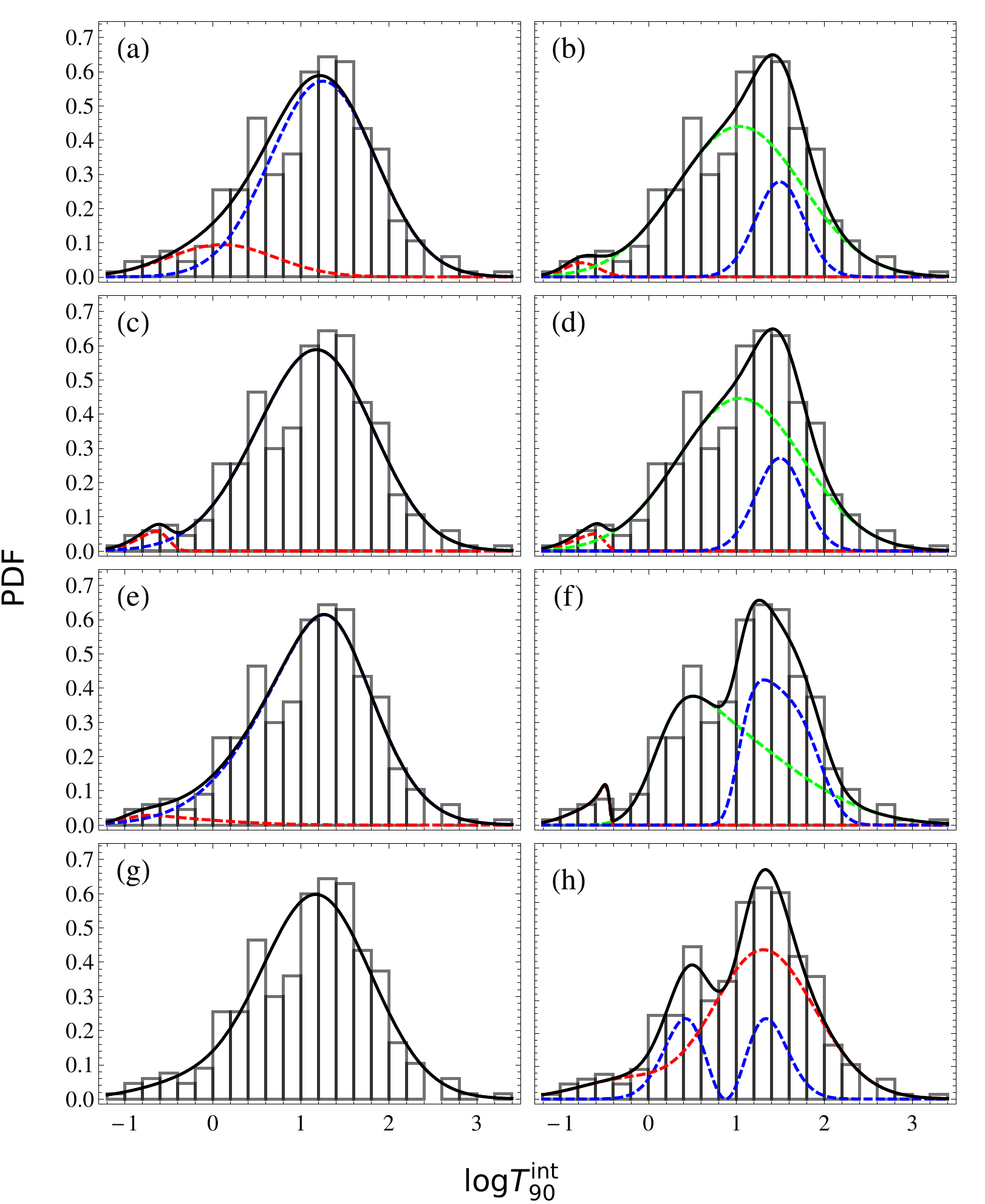}
\caption{The same as Fig.~\ref{fig3}, but for $\log T^{\rm int}_{90}$ and for {\it Swift} GRBs.}
\label{fig9}
\end{figure}

\begin{figure}
\includegraphics[width=\columnwidth]{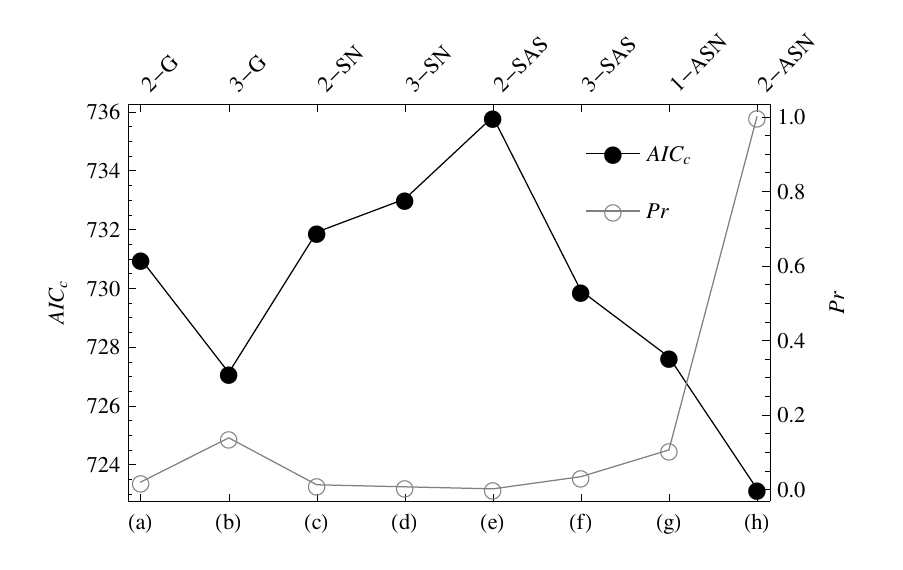}
\caption{The same as Fig.~\ref{fig4}, but for intrinsic durations and the {\it Swift} GRBs.}
\label{fig10}
\end{figure}

\section{Discussion}\label{disc}

Generally, inferring an existence or lack of thereof, based on statistical evidence, must be done with care. Having samples with limited size adds difficulty to such an assessment, as in small samples there is more room for statistical fluctuations that might obscure the global picture. Previous researches, cited in Sect.~\ref{intro}, mostly imply that a 3-G fit is a better descriptive model than a 2-G. Nevertheless, the fits achieved were bimodal, indicating the presence of only two GRB classes \citep{Tarnopolski}. A remarkable exception was the BATSE 3B dataset \citep{horvath98}, where the third peak had a negligible probability of $10^{-4}$ to be a chance occurence. It turnt out, however, that a bigger dataset obtained by the same instrument did not reveal its presence anymore \citep{horvath02}. Examining the observed, instead of intrinsic, durations might also cast doubts on the reality of the observed phenomenon. Having that in mind, it is tempting to state that the intermediate GRB class is unlikely to be a real class based on the analysis of 408 GRBs with known both $T_{90}$ and redshift. This statement could be justified with the results presented in Sect.~\ref{res1}, where the two best models to describe the $\log T^{\rm obs}_{90}$ distribution were trimodal, but after  moving to the rest frame, the most plausible description was provided with a conventional 2-G. It may appear that the intrinsic durations should trace the physical context of the GRBs more appropriately.

On the other hand, the GRB characteristics are not only sample-dependent, as showed above, but also detector-dependent \citep[e.g.,][]{horvath06,horvath09,huja,ripa,horvath10,bromberg2,zitouni,Tarnopolski2,Tarnopolski3}. Thereofore, lacking a dataset numerous enough for the statistics to provide a convincing proof, one may only claim evidence, or its lack, in a specific sample under consideration (see also \citealt{Tarnopolski,Tarnopolski2}). To get rid of the detector-dependency, only 334 GRBs as detected by {\it Swift} were examined. The outcome of this analysis, shown in Sect.~\ref{res2}, is surprisingly inconsistent with the one from a bigger sample in Sect.~\ref{res1}, being at the same time consistent with previous analyses performed on a bigger sample of {\it Swift} GRBs---the obtained fits are all uni- or bimodal, and the one with the lowest $AIC_c$ is a bimodal 3-G; the next best fits were a unimodal 1-ASN and a bimodal 2-SN. Both yield strong evidence in their favor, so it is not possible to unambiguously infer the number of components, or even the modality of the {\it Swift} sample\footnote{This, combined with the fact that the number of GRBs with measured redshift is relatively low, may be due to the fact that {\it Swift} is more sensitive in soft bands than BATSE was, hence its dataset has a lower fraction of short GRBs}.

After moving to the rest frame, the problems are not solved, especially that the best fit now is a 2-ASN. The problem with this distribution is that it consists of a bimodal component [see Fig.~\ref{fig9}~(h)], with locations of its peaks in agreement with the groups of short and long GRBs. It seems like the role of the second component here is to merely adjust the height of the fit. The second best fit, a bimodal 3-G, has a relative probability of 13.9\%. While this is a statistically valid result, meaning that among the exmined distributions the 2-ASN is best balanced between the goodness of fit and the number of parameters, from the physical point of view, regarding the knowledge about GRBs, this result is an unrealistic one, as the short and long GRBs are known to stem from different progenitors, mergers and collapsars, respectively. Even after dismissing the 2-ASN, differentiating between a 3-G and 1-ASN is not possible in the framework of the $AIC_c$, as these two models yield $\Delta_i=0.535$. Hence, the currently available redshift distribution unfortunately does not allow to infer the existence of the intermediate GRBs class reliably, likely due to the smallness of the sample.

\section{Conclusions}\label{conc}

The research conducted so far on different samples of GRB duration distributions indicate that a 3-G follows the data better than a 2-G. However, even with three components, the fitted distribution is usually bimodal, implying two physical classes. Because a two-component mixture of skewed distributions was shown to be a statistically better fit, in case of {\it Fermi}/GBM observations, than the commonly applied 3-G \citep{Tarnopolski2}, in this paper the same approach was undertaken to investigate the modality and goodness of fit in case of GRBs with measured both redshift and duration. The reason for this is that in the rest frame the effects of cosmological factors are mostly elliminated, hence it is expected that it will provide an insight into the properties of GRBs.

It was found that in a sample of 408 GRBs with known redshift, the best fits---3-G and 3-SN---are trimodal (in the sense of having three local maxima), but after moving to the rest frame, a (unimodal) 2-G yielded considerably stronger support than any other examined distribution. However, this sample is dominated by detections made by {\it Swift}/BAT (334 events, $\approx 82\%$ of the total number of GRBs with measured $z$), and hence this finding might be affected by the fact that GRB properties are detector-dependent. Therefore, the {\it Swift}/BAT subsample was also examined, and it turnt out that it is not possible to reliably infer the best fit within the chosen information-theoretic framework ($AIC_c$ in this work). This may be caused by the smallness of the sample, and so the solution is to, hopefully, repeat the analysis in the future on a wider GRB sample (see also \citealt{zhang3}). Because the mathematical model of the observed as well as intrinsic durations is still lacking, the physical interpretation of the results obtained herein is limitted. The distribution of intrinsic durations, being systematically shifted towards shorter values, while may be believed to trace the properties of GRB population more accurately, is also affected by statistical fluctuations. Considering the {\it Swift} subsample, the distributions are strongly dominated by long GRBs, what might cause introduction of biases in the analysis undertaken.

\acknowledgments
The author acknowledges support in a form of a special scholarship of Marian Smoluchowski Scientific Consortium Matter-Energy-Future from KNOW funding, grant number KNOW/48/SS/PC/2015.

\end{document}